\documentclass[final,authoryear,3p,times,10pt]{elsarticle}
\usepackage{amssymb,mathtools,amsmath,lscape}
\usepackage{multirow,setspace,times,graphicx,color,rotating,subfigure,url}
\usepackage{booktabs}
\usepackage{multicol}
\usepackage{threeparttable}
\usepackage{longtable,pdflscape}

\usepackage{pgfplots}
\pgfplotsset{compat=newest}
\usepgfplotslibrary{groupplots}
\usepgfplotslibrary{dateplot}

\begin{document}

\begin{frontmatter}
\title{Sector connectedness in the Chinese stock markets}


\author[DF&RCE]{Ying-Ying Shen}
\author[DF&RCE]{Zhi-Qiang Jiang \corref{cor}} \ead{zqjiang@ecust.edu.cn} %
\author[DF&RCE]{Jun-Chao Ma} 
\author[HNU]{Gang-Jin Wang \corref{cor}} \ead{wanggangjin@hnu.edu.cn}
\author[DF&RCE]{Wei-Xing Zhou} \ead{wxzhou@ecust.edu.cn} %
\cortext[cor]{Corresponding authors.}

\address[DF&RCE]{School of Business and Research Center for Econophysics, East China University of Science and Technology, Shanghai 200237, China}
\address[HNU]{Business School and Center for Finance and Investment Management, Hunan University, Changsha 410082, China} %

\begin{abstract}
Uncovering the risk transmitting path within economic sectors in China is crucial for understanding the stability of the Chinese economic system, especially under the current situation of the China-US trade conflicts. In this paper, we try to uncover the risk spreading channels by means of volatility spillovers within the Chinese sectors using stock market data. By applying the generalized variance decomposition framework based on the VAR model and the rolling window approach,  a set of connectedness matrices is obtained to reveal the overall and dynamic spillovers within sectors. It is found that 17 sectors (mechanical equipment, electrical equipment, utilities, and so on) are risk transmitters and 11 sectors (national defence, bank, non-bank finance, and so on) are risk takers during the whole period. During the periods with the extreme risk events (the global financial crisis, the Chinese interbank liquidity crisis, the Chinese stock market plunge, and the China-US trade war), we observe that the connectedness measures significantly increase and the financial sectors play a buffer role in stabilizing the economic system. The robust tests suggest that our results are not sensitive to the changes of model parameters. Our results not only uncover the spillover effects within the Chinese sectors, but also highlight the deep understanding of the risk contagion patterns in the Chinese stock markets.

\end{abstract}


\end{frontmatter}

\section{Introduction}

The China-US trade war has severe impacts on the Chinese stock markets. During the period from 2008 to 2019, the Chinese stock markets exhibit extremely volatile, that the Shanghai Composite Index lost more than 1,000 points in 2018 (from the maximum 3587 to the minimum 2440) and followed by a very hard recovery of 500 points in 2019. It is observed that the releasing of information about adding tariffs on specific Chinese products undermines the confidence of investors and negatively shocks the markets. Naturally, such negative shocks may have direct impacts on the sectors which export products on the US tariff lists and then transmit to the other sectors having direct and indirect connections to the targeted sectors, mediated by global value chain interdependencies \citep{Egger-Zhu-2019-XXX}. Because of the recent global financial crises, there are a plenty of empirical studies focusing on the risk contagions in different financial markets \citep{Jung-Maderitsch-2014-JBF}, in different financial institutions or assets \citep{McKibbin-Martin-Tang-2014-JBF, Elyasiani-Kalotychou-Staikouras-Zhao-2015-JFSR}, and in different countries \citep{Corradi-Distaso-Fernandes-2012-JEC, Cotter-Suurlaht-2019-EJF, Kenourgios-Samitas-Paltalidis-2011-JIFMIM}. However, very few efforts have been put on uncovering risk transmission between different economic sectors within China, which is especially crucial for understanding the stability of the Chinese economic system under the situation of the China-US trade conflicts.

Our goal here is to analyze the spillover dynamics within a set of Chinese sectors. Following \cite{Egger-Zhu-2019-XXX}, our analysis is performed on the daily sector indexes from the Chinese stock markets, which allows us to detect immediate responses to new events. A measure of connectedness or spillover proposed by \cite{Diebold-Yilmaz-2012-IJF, Diebold-Yilmaz-2014-JE} is adopted to explore the direct and net spillovers across sectors. The connectedness measure is constructed based on the concept of forecast error variance decomposition and is independence of the variable orders under the framework of vector autoregression (VAR) \citep{Diebold-Yilmaz-2012-IJF, Diebold-Yilmaz-2014-JE}. This method also allows us to examine the time-varying dynamics of the connectedness index by means of a rolling-window approach. With the same method, \cite{Fonseca-Ignatieva-2018-AE} study volatility spillover effects among ten CDS sector indexes in the US markets and \cite{Collet-Ielpo-2018-JBF} investigate spillovers within eight sectors in the US credit markets.  

The contributions of this paper are listed as follows. First, we extend the empirical literature by taking into account the spillover dynamics within economic sectors, especially paying attentions on the risk contagion behaviors around the extreme risk events, which complements the existing studies on revealing the spillovers between oil markets and sectors \citep{Arouri-Jouini-Nguyen-2011-JIMF, Mensi-Hammoudeh-Jarrah-Sensoy-Kang-2017-EE, Ahmad-Mishra-Daly-2018-IRFA, Wang-Wang-2019-EE} and between institutions within financial sectors \citep{Wang-Xie-Zhao-Jiang-2018-JIFMIM, Cotter-Suurlaht-2019-EJF, Li-Yao-Li-Zhu-2019-IJFE, Peltonen-Rancan-Sarlin-2019-IJFE}. Second, differing from the researches on the sector connectedness in the US CDS markets \citep{Fonseca-Ignatieva-2018-AE} and in the US credit markets \citep{Collet-Ielpo-2018-JBF}, our work focuses on the sector connectedness within the Chinese stock markets. In our analysis, our data set is comprised of 28 different sectors, which is much larger than the sector number in \cite{Fonseca-Ignatieva-2018-AE} and \cite{Collet-Ielpo-2018-JBF}. Our data set spanning a period from 4 January 2000 to 31 December 2019 not only covers the recent global financial crisises, but also includes the China-US trade conflicts, which allows us to compare the risk contagion behaviors triggered by the two different events in the Chinese economic sectors. Third, \cite{Egger-Zhu-2019-XXX} quantitatively estimate the impact of increasing 1\% tariff by the US on the targeted sectors and indirect sectors in China. However, the spreading processes from the targeted sectors to the other sectors are not clear. We thus fill this gap by constructing a connectedness network based on the connecteddness matrix given by \cite{Diebold-Yilmaz-2014-JE} to detect the risk diffusing channels under the situation of the China-US trade conflicts.

This paper is organized as follows. A brief literature review is presented in Section 2. Section 3 introduces the data of sector indexes, the definition of volatility, and the estimation of connenctedness measure. Sections 4 presents the overview of the connectedness measure. Sections 5 and 6 give the evolution of the total connectedness and the pairwise directional connectedness. Section 7 discusses the net pairwise connectedness network in the subperiods having typical extreme risk events. The robust tests on different model parameters are presented in Section 8. Section 9 concludes.

\section{Literature review}

Due to the high frequent occurrence of financial crises in recent decades, the research topics related to contagion \citep{Karolyi-Stulz-1996-JF, Allen-Gale-2000-JPE, Pericoli-Sbracia-2003-JES, Forbes-Rigobon-2002-JF} and interdependence \citep{Broner-Gelos-Reinhart-2006-JIE} have been received considerable research interests. Contagion usually refers to that the negative event (crisis) occurred in one market or region has the amplification of spillover effects on other markets or regions \citep{Karolyi-Stulz-1996-JF, Allen-Gale-2000-JPE, Forbes-Rigobon-2002-JF}. Differing from contagions, interdependence corresponds to the cross-market linkage that there is no major change in market fundamentals \citep{Forbes-Rigobon-2002-JF}. \cite{Pericoli-Sbracia-2003-JES} also distinguish contagion and interdependence through clarifying the propagating channels of negative shocks. Although there are differences between contagion and interdependence, the index of financial interdependence, namely the spillover index, can also be used to quantitatively present the market co-movements, in particularly, the evolution of risk contagion during crisis periods \citep{Broner-Gelos-Reinhart-2006-JIE}. 

\subsection{Contagion}

There are two main strands of literature on financial contagion. One is to capture the pairwise correlations between institutions or markets based on correlation measures \citep{Baele-2005-JFQA, Chiang-Jeon-Li-2007-JIMF, Teply-Kvapilikova-2017-NAJEF, Wang-Xie-Lin-Stanley-2017-FRL}. The widely used methods include the GARCH models \citep{Engle-2002-JBES}, the Granger causality models \citep{Sander-Kleimeier-2003-JIFMIM}, the copula models \citep{Patton-2006-IER, Ning-2010-JIMF}, the CoVaR \citep{Adrian-Brunnermeier-2011-NBER}, the quantile regression analysis \citep{Nusair-Olson-2019-EE}, and so on. The multivariate GARCH models are initially proposed to study the spillover between different assets. However, such model has difficulties in parameter estimation. Thus, a DCC-GARCH model is tailored for the high-dimensional system \citep{Engle-2002-JBES}, in which spillover is characterized by dynamic conditional correlations, and successfully applied to analyze up to 100 assets \citep{Engle-Sheppard-2001-NBER}. In empirical analysis, the number of assets is still limited by the computational and presentational difficulties for the DCC-GARCH model. An updated model named dynamic equicorrelation (DECO) is proposed to tackle such drawback \citep{Engle-Kelly-2012-JBES}. As the GARCH models are unable to capture the direction of contagion, the Granger test is applied to uncover the direct risk spreading patterns. Using the Granger causality tests, \cite{Sander-Kleimeier-2003-JIFMIM} provide a directional contagion pattern at the regional level during Asian crisis and \cite{Kalbaska-Gatkowski-2012-JEBO} confirm the existence of contagion effects within France, Germany, the United Kingdom and other countries through direct causality networks. Both GARCH models and Granger tests have the drawback of underestimating tail risk and tail dependence, thus more sophisticate models are invented, for example the copula and CoVaR. \cite{Wen-Wei-Huang-2012-EE} apply the copula and CoVaR to study the contagion effect between energy and stock markets during the recent financial crises and observe a significant characteristic of increasing tail dependence which is a manifestation of contagion.

The other is to uncover the path of financial contagions from the perspective of complex networks. Similar to the rumour spreading on social networks, financial networks shape the risk contagion, thus play an important role in stabilizing the financial system. \cite{Gai-Kapadia-2010} analytically reveal that the changes of network structure have direct influences on the probability of risk contagion. \cite{Huang-Zhuang-Yao-Uryasev-2016-PA} investigate the potential effects of local structures of financial networks on the stability of financial system and argue that financial nodes with greater strength, centrality, tightness, and clustering coefficients have greater contribution to systemic risks. This means that high interwoven of local structure in financial networks has an amplifying effect on local risk. However, such amplifying effect is not always true. When the negative impacts are small enough, financial networks with dense connections can enhance the stability of financial system \citep{Acemoglu-Ozdaglar-Tahbaz-Salehi-2015-AER}. 

\subsection{Interdependence}

The fluctuations in one market or one region are found to have influences on the other markets and regions. \cite{Hamao-Masulis-Ng-Uryasev-1990-RFS} find that for the markets with different trading time, the shocks happened in one market may have impacts on the markets which open in the following. \cite{Melvin-Melvin-2003-RES} provide evidence that volatilities of stock markets are originated from those of exchange markets. Oil shocks are considered as an important source of risks and have great impacts on financial markets, which inspires the researches on revealing the causality, interdependence, and spillovers between oil markets and stock markets. By applying the Granger causality tests on the prices of oil and renewable energy company stocks, \cite{Henriques-Sadorsky-2008-EE} find that the technology stocks and oil are Granger causes to the renewable energy companies.  By employing the multivariate GARCH (MGARCH) model to analyze the volatility spillover effects within oil prices, stock prices of technology, and stock prices of clean energy companies, \cite{Sadorsky-2012-EE} found that the correlation between the clean energy company stock prices and the technology company stock prices is much stronger that between the oil prices and the prices of clean energy companies. 

\cite{Diebold-Yilmaz-2012-IJF, Diebold-Yilmaz-2014-JE} propose an econometric method (abbreviated as DY method hereafter) to capture the dynamic spillovers within a set of representative financial variables (like returns or volatilities) associated with countries, markets, institutions or companies, and so on. Intriguingly, this method is extremely useful to reveal the evolution of spillovers when it is combined with a rolling window approach, which leads to the wide applications on risk spillover analysis \citep{Fernandez-Rodriguez-Gomez-Puig-Sosvilla-Rivero-2016-JIFMIM, Yang-Zhou-2017-MS, Barunik-Kocenda-Vacha-2017-JIMF, Dahl-Jonsson-2018-JCM, Yoon-Mamun-Uddin-Kang-2019-NAJEF}. For example, DY method is applied to investigate the dynamic spillover effects among commodity futures (crude oil, precious metal, and agricultural products) \citep{Kang-McIver-Yoon-2017-EE} and to detect the asymmetric volatility connectedness in exchange markets during the periods of financial crises \citep{Barunik-Kocenda-Vacha-2017-JIMF}. Furthermore, DY method is also employed to assess volatility spillovers in the seafood markets \citep{Dahl-Jonsson-2018-JCM}. The connectedness matrix generated by DY method is also used to construct risk spillover networks. \cite{Yang-Zhou-2017-MS} identify networks of volatility spillovers and investigate time-varying spillover among the U.S. Treasury bonds, global stock indexes, and commodities. Besides the markets mentioned in \cite{Yang-Zhou-2017-MS}, \cite{Yoon-Mamun-Uddin-Kang-2019-NAJEF} also add the currency markets to investigate the risk spillover within the markets of bond, stock, commodity, and exchange. Both studies report that the US stock market is the greatest single risk contributor in the global financial markets \citep{Yang-Zhou-2017-MS, Yoon-Mamun-Uddin-Kang-2019-NAJEF}. By building the connectedness network of EMU countries using the sovereign bond data, \cite{Fernandez-Rodriguez-Gomez-Puig-Sosvilla-Rivero-2016-JIFMIM} identify the countries whether they locate in the network core or network periphery and find that the risk transmits from the core to the periphery. More importantly, such connectedness network can be further used of bank supervision and financial stability monitoring \citep{Wang-Xie-Zhao-Jiang-2018-JIFMIM, Hale-Lopez-2019-JEC}.

In contrast to the widely application of DY method in empirical studies, little attention has been paid to the spillover effects within sectors. A handful papers report the credit risk transmission within CDS sectors in the credit markets \citep{Collet-Ielpo-2018-JBF, Fonseca-Ignatieva-2018-AE, Shahzad-Bouri-Arreola-Roubaud-Bekiros-2019-AE}. Particularly, \cite{Collet-Ielpo-2018-JBF} investigate the cross-sector volatility spillovers in the US credit market and find that the sectors exhibit very high spillover effects and the risk contributors are comprised of insurance, commodity, and energy sectors. By performing very similar analysis on the sectors in stock markets, \cite{Nguyen-Nguyen-Tan-2018-SSRN} move one step further to construct a complete tail risk connectedness network for the entire US industrial system and find that the tail risk dependence is mainly driven by the trade flow within sectors. Practically, examining the sectoral spillover effects in stock markets is helpful to find potential sector-based hedging opportunities for investors. As no market is isolated in the global financial markets, the sectors in one market could receive risk shocks from other markets. By investigating the dynamic risk spillovers across two major commodity markets (crude oil and gold), the aggregate Dow Jones Islamic (DJIM) index, and ten stock sectors, \cite{Mensi-Hammoudeh-Jarrah-Sensoy-Kang-2017-EE} find that the oil and gold markets and the sectors of energy, finance, technology, and telecommunication are net risk takers, receiving risks from the DJIM index and the other sectors. \cite{Yu-Du-Fang-Yan-2018-IREF} explore the risk contribution of crude oil markets to the sectors in the US stock markets and find that the crude oil contributes the greatest risk to the energy sector and transmits the least risk to the consumer staple sector. 

\section{Method and data}

\subsection{Data}

Our analysis is performed on 28 different sector indices in the Chinese stock markets issued by Shenyin \& Wanguo Securities Co., Ltd. (http://www.swsresearch.com). The sector names and their abbreviations are listed as follows: Agriculture and forestry (A\&F), Mining (Mining), Chemical (Chem), Steel (Steel), Non-ferrous metals (NFMet), Electronic (Elec), Household appliances (HApp), Food and drink (F\&D), Textile and apparel (T\&A), Light manufacturing (LMF), Biotechnology (Biotech), Utilities (Util), Transportation (Trans), Real estate (REst), Commercial trade (ComT), Leisure and services (L\&S), Composite (Cps), Building materials (BM), Building and decoration (B\&D), Electrical equipment (ElecE), National defence (ND), Computer (Cpt), Media (Media), Communications (Comm), Bank (Bank), Non-bank financial (NBF), Automobile (Auto), and Mechanical equipment (MechE). Our data span a period from 4 January 2000 to 31 December 2019 and on each trading day the open, high, low, and close indexes are recorded. Note that there is an index adjustment on 21 February 2014 for each index, which leads to an extreme volatility on that day. We thus remove the volatility on 21 February 2014 for all indexes. 

To investigate the volatility spillover effects, we first define the daily volatility for each index following \cite{Garman-Klass-1980-JB},
\begin{equation}
V_{it}^{GK}=0.511(H_{it}-L_{it})^2-0.019\left[(C_{it}-O_{it})(H_{it}+L_{it}-2O_{it})-2(H_{it}-O_{it})(L_{it}-O_{it}) \right]-0.383(C_{it}-O_{it})^2,
\label{Eq:Volatility}
\end{equation}
where $H_{it}$, $L_{it}$, $O_{it}$, and $C_{it}$ are the natural logarithm of high, low, open, and close values of index $i$ on day $t$. For each index, once we obtain its volatility, the corresponding mean, median, maximum, minimum, standard deviation, skewness, kurtosis, and ADF statistics are also estimated. 

\begin{table}[t]
\setlength\tabcolsep{5pt}
\centering 
\small
\caption{\label{Tb:DS} Descriptive statistics of the sector volatilities. This table reports sector name, abbr. of sector name, sector code, as well as the mean, median, max, min, std, skewness (skew.), kurtosis (kurt.), ADF test statistic (ADF) of the volatilities. The ADF tests are performed at the lag of 2 and the results are all significant at the level of 1\%.}
\begin{tabular}{llcr@{.}lr@{.}lr@{.}lr@{.}lr@{.}lr@{.}lr@{.}lr@{.}l}
\toprule
\multirow{2}{*}{name} & \multirow{2}{*}{abbr} & \multirow{2}{*}{code} & \multicolumn{2}{c}{mean} &  \multicolumn{2}{c}{median} 
& \multicolumn{2}{c}{max} &  \multicolumn{2}{c}{min} 
& \multicolumn{2}{c}{std} & \multicolumn{2}{c}{\multirow{2}{*}{skew.}} 
& \multicolumn{2}{c}{\multirow{2}{*}{kurt.}} & \multicolumn{2}{c}{\multirow{2}{*}{ADF}}\\
& & & \multicolumn{2}{c}{$(\times10^{-4})$} &\multicolumn{2}{c}{$(\times10^{-4})$} &\multicolumn{2}{c}{$(\times10^{-4})$} & \multicolumn{2}{c}{$(\times10^{-8})$} & \multicolumn{2}{c}{$(\times10^{-4})$} & \multicolumn{2}{c}{} & \multicolumn{2}{c}{} & \multicolumn{2}{c}{} \\
\midrule
Agriculture and forestry & A\&F & 801010  & 3&07  & 1&66  & 79&28  & 159&16  & 4&57  & 5&25  & 50&47  & -15&04$^{***}$ \\
Mining & Mining & 801020  & 3&29  & 1&72  & 89&17  & 157&44  & 4&90  & 4&85  & 43&72  & -15&40$^{***}$ \\
Chemical & Chem & 801030  & 2&51  & 1&33  & 79&71  & 132&99  & 3&82  & 5&73  & 65&03  & -16&01$^{***}$ \\
Steel & Steel &801040  & 2&90  & 1&55  & 93&02  & 293&48  & 4&40  & 5&66  & 64&10  & -16&53$^{***}$ \\
Non-ferrous metals & NFMet & 801050  & 3&35  & 1&90  & 65&05  & 454&31  & 4&59  & 4&35  & 34&08  & -15&19$^{***}$ \\
Electronic & Elec & 801080  & 3&06  & 1&75  & 84&91  & 217&92  & 4&28  & 5&14  & 52&67  & -15&43$^{***}$ \\
Household appliances & HApp & 801110  & 2&70  & 1&62  & 60&35  & 50&17  & 3&81  & 5&49  & 51&88  & -15&44$^{***}$ \\
Food and drink & F\&D & 801120  & 2&33  & 1&35  & 83&81  & 382&48  & 3&53  & 7&38  & 107&60  & -15&89$^{***}$ \\
Textile and apparel & T\&A & 801130  & 2&64  & 1&33  & 118&41  & 60&58  & 4&38  & 7&19  & 121&44  & -16&97$^{***}$ \\
Light manufacturing & LMF & 801140  & 2&55  & 1&32  & 76&40  & 0&48  & 3&98  & 5&66  & 58&61  & -16&15$^{***}$ \\
Biotechnology & Biotech & 801150  & 2&45  & 1&27  & 100&53  & 6&48  & 3&97  & 7&38  & 114&06  & -16&50$^{***}$ \\
Utilities & Util & 801160  & 2&33  & 1&07  & 72&14  & 54&13  & 3&99  & 5&97  & 62&47  & -15&86$^{***}$ \\
Transportation & Trans & 801170  & 2&36  & 1&16  & 61&44  & 12&07  & 3&90  & 5&86  & 58&10  & -15&75$^{***}$ \\
Real estate & REst & 801180  & 3&16  & 1&79  & 82&74  & 314&18  & 4&57  & 5&31  & 53&36  & -15&53$^{***}$ \\
Commercial trade & ComT & 801200  & 2&58  & 1&29  & 112&08  & 19&74  & 4&37  & 8&27  & 137&24  & -16&48$^{***}$ \\
Leisure and services & L\&S & 801210  & 3&20  & 1&73  & 87&85  & 14&63  & 4&74  & 5&10  & 49&71  & -15&20$^{***}$ \\
Composite & Cps & 801230  & 2&93  & 1&56  & 94&19  & 14&56  & 4&22  & 5&38  & 65&50  & -14&73$^{***}$ \\
Building materials & BM & 801710  & 6&18  & 4&25  & 87&36  & 299&37  & 6&83  & 2&72  & 15&87  & -9&81$^{***}$ \\
Building and decoration & B\&D & 801720  & 5&09  & 3&32  & 71&38  & 261&65  & 6&10  & 3&24  & 19&37  & -11&07$^{***}$ \\
Electrical equipment & ElecE & 801730  & 5&83  & 3&98  & 71&63  & 185&42  & 6&55  & 2&80  & 16&33  & -10&29$^{***}$ \\
National defense & ND & 801740  & 7&18  & 4&82  & 87&49  & 196&69  & 7&87  & 3&16  & 20&30  & -11&76$^{***}$ \\
Computer & Cpt & 801750  & 6&51  & 4&84  & 117&44  & 367&35  & 6&65  & 3&33  & 28&53  & -11&64$^{***}$ \\
Media & Media & 801760  & 6&76  & 4&83  & 81&26  & 36&79  & 6&96  & 2&60  & 15&46  & -10&66$^{***}$ \\
Communications & Comm & 801770  & 5&23  & 3&63  & 93&81  & 52&37  & 5&88  & 3&93  & 31&29  & -13&18$^{***}$ \\
Bank & Bank & 801780  & 3&58  & 1&89  & 72&84  & 4&86  & 5&21  & 4&48  & 35&09  & -14&97$^{***}$ \\
Non-bank financial & NBF & 801790  & 6&16  & 3&78  & 131&81  & 26&22  & 7&62  & 4&07  & 34&56  & -13&16$^{***}$ \\
Automobile & Auto & 801880  & 5&92  & 4&28  & 80&47  & 6&96  & 6&44  & 3&07  & 20&35  & -10&85$^{***}$ \\
Mechanical equipment & MechE & 801890  & 6&00  & 4&51  & 63&03  & 338&19  & 6&42  & 2&69  & 14&73  & -10&09$^{***}$ \\
\bottomrule
\end{tabular}
\end{table}

Table \ref{Tb:DS} provides the summary statistics for each sector volatility during the entire period. We can find that for each volatility the mean value is larger than its median value, suggesting that the tail is on the right, in accordance with the positive skewness value. The kurtosis is much greater than 3 and the skewness is positive, implying that the sector volatility exhibits the characteristics of leptokurtosis and fat-tail. Furthermore, at the lag order of 2 and the significant level of 1\%, the ADF statistics are significant, indicating the rejection of having a unit root in volatility.

\subsection{Connectedness}

\cite{Diebold-Yilmaz-2014-JE} propose a connectedness measure to investigate spillover effects among different financial institutions in a framework of variance decomposition . The connectedness measure is constructed as follows. Let's consider a vector autoregressive model (VAR) with $N$ variables for a covariance stationary process \citep{Diebold-Yilmaz-2009-EJ, Diebold-Yilmaz-2012-IJF, Diebold-Yilmaz-2014-JE},
\begin{equation}
Y_t = \sum_{i=1}^{p} \Phi_{t-i} Y_{t-i} + \varepsilon_t, ~~ \varepsilon \sim N(0, \Sigma ), 
\label{Eq:VAR:Yt}
\end{equation}
where $Y_t = (y_{1t}, y_{2t}, ..., y_{Nt})'$ is a $N \times 1$ vector of endogenous variables at time $t$ ($t = 1, 2, ..., T$), $\Phi_i$ ($i = 1, 2, ..., p$) is a parameter matrix, $\varepsilon$ is a vector of i.i.d. white noises, and $p$ is the lag order of VAR. $Y_t$ can also be written in a moving average form,
\begin{equation}
Y_t = \sum_{i=0}^{\infty} A_{i}\varepsilon_{t-i},
\label{Eq:MA:Yt}
\end{equation}
where $A_i$ is a $N \times N$ coefficient matrix and obtained through a recursive way $A_i = \Phi_{1}A_{i-1} + \Phi_{2}A_{i-2} + \cdots + \Phi_{p}A_{i-p}$ with $A_0$ being an identity matrix and $A_i = 0$ for $i<0$. In the framework of variance decomposition, we can decompose the forecast error variance of each variable into parts that are attributed to various shocks from other variables. 

We denote $D^{gH}=\left[d_{ij}^{gH}\right]$ as the $H$-step generalized variance decomposition matrix.  $H$ is the predictive horizon.  The element $d_{ij}^{gH}$ corresponds to the fraction of sector $i$'s $H$-step-ahead generalized forecast error variance due to the shocks from sector $j$ and is calculated as follows, 
\begin{equation}
d_{ij}^{gH} = \sigma_{jj}^{-1} \frac{ \sum_{h=0}^{H-1} \left(e_i'A_h \Sigma e_j \right)^2 } { \sum_{h=0}^{H-1} {\left( e_i' A_h \Sigma A_h' e_i \right )}},
\label{Eq:Dgh}
\end{equation}
where $\sigma_{jj}$ is the $j$-th diagonal element of covariance matrix and $e_j$ is a selection vector with $e_j(k) = 1$ if $k = j$ and zero otherwise. $\Sigma$ is the covariance matrix of the shock vectors in the non-orthogonal VAR. Due to the lack of orthogonality in the generalized variance decomposition framework, the sum of the forecast error variance contributions is not ensured to be 1, such that $\sum_{j=1}^{N} d_{ij}^{gH} \neq 1$. We thus define a normalized matrix $\tilde{D}^{gH} = \left[\tilde{d}_{ij}^{gH}\right]$, which is calculated by normalizing the matrix $D^{gH}$ for each row,
\begin{equation}
\tilde{d}_{ij}^{gH} = \frac{ d_{ij}^{gH} }{ \sum_{j=1}^{N}{d_{ij}^{gH}} }.
\label{Eq:Dgh:norm:row}
\end{equation}
One can have $\sum_{j=1}^{N} {\tilde{d}_{ij}^{gH}} = 1$ and $\sum_{i,j=1}^{N} {\tilde{d}_{ij}^{gH}} = N$. The normalized matrix $\tilde{D}^{gH}$ is nothing but the adjacent matrix of volatility connectedness network.

To capture the direction of connectedness, we define the directional connectedness from sector $j$ (respectively, $i$) to sector $i$ ($j$) as $C_{i \leftarrow j}^H = \tilde{d}_{ij}^{gH}$ ($C_{j\leftarrow i}^H = \tilde{d}_{ji}^{gH}$). Usually, $C_{j \leftarrow i}^H$ does not equal to $C_{i\leftarrow j}^H$. We further denote the directional volatility spillovers received by sector $i$ from all other sectors as from-connectedness, 
\begin{equation}
C_{i\leftarrow{\cdot}}^{H} = \frac{\sum_{j=1, j\neq i}^{N} \tilde{d}_{ij}^{gH}} {\sum_{j=1}^{N} \tilde{d}_{ij}^{gH}},
\label{Eq:from:con}
\end{equation}
and the directional volatility spillovers transmitted by sector $i$ to all other sectors as to-connectedness,
\begin{equation}
C_{{\cdot}\leftarrow i}^{H} = \frac{ \sum_{j=1, j\neq i}^{N} {\tilde{d}_{ji}^{gH}}} {\sum_{j=1}^{N}{\tilde{d}_{ji}^{gH}}}.
\label{Eq:to:con}
\end{equation}
The difference between to-connectedness and from-connectedness is defined as net-connectedness (from sector $i$ to all other sectors),
\begin{equation}
C_i^H = C_{\cdot \leftarrow i}^H - C_{i \leftarrow \cdot}^H.
\label{Eq:net:con}
\end{equation}
$C_i^H$ measures the net spillover effect. The three connectedness measures are able to provide the information that each sector receives (transmits) spillovers from (to) the other sectors. Furthermore, we introduce the total connectedness to measure the contribution of directional connectedness of all sectors, which is given by the mean of the off-diagonal elements in $D^H$,
\begin{equation}
C_T^H = \frac{1}{N} \sum_{i,j=1,j \neq i}^{N} {\tilde{d}_{ij}^{gH}}.
\label{Eq:net:total}
\end{equation}
Note that $C_T^H$ locates in the range of $[0\%, 100\%]$. The relationships of the mentioned connectedness measures are also illustrated in Table~\ref{Tb:connectedness}.

\begin{table}[htb]
\setlength\tabcolsep{7pt}
\centering 
\small
\caption{Definition of the connectedness measures. $V_i$ represents the volatility of the $i$-th sector. The $\tilde{d}_{ij}^{gH}$ is the fraction of industry $i$'s $H$-step-ahead generalized forecast error variance due to the shocks from sector $j$. $C_{i\leftarrow{\cdot}}^{H}$ and $C_{{\cdot}\leftarrow i}^{H}$ stand for from-connectedness and to-connectedness, respectively. }
\label{Tb:connectedness}
\vskip 2.5mm
\begin{tabular}{cccrlcc}
\toprule
{} & $V_1$ & $V_2$ & \multicolumn{2}{c}{$\cdots \cdots$} & $V_N$ & $C_{i\leftarrow{\cdot}}^{H}$ \\
\midrule
$V_1$ & $\tilde{d}_{11}^{gH}$ & $\tilde{d}_{12}^{gH}$ & \multicolumn{2}{c}{$\cdots \cdots$} & $\tilde{d}_{1N}^{gH}$ & $\sum_{j=1}^{N}{\tilde{d}_{1j}^{gH}},j\neq1$\\
$V_2$ & $\tilde{d}_{21}^{gH}$ & $\tilde{d}_{22}^{gH}$ & \multicolumn{2}{c}{$\cdots \cdots$} & $\tilde{d}_{2N}^{gH}$ & $\sum_{j=1}^{N}{\tilde{d}_{2j}^{gH}},j\neq2$\\
$\vdots$ & $\vdots$ & $\vdots$ & \multicolumn{2}{c}{$\ddots$} & $\vdots$ & $\vdots$\\
$V_N$ & $\tilde{d}_{N1}^{gH}$ & $\tilde{d}_{N2}^{gH}$ & \multicolumn{2}{c}{$\cdots \cdots$} & $\tilde{d}_{NN}^{gH}$ & $\sum_{j=1}^{N}{\tilde{d}_{Nj}^{gH}},j\neq N$\\
\midrule
$C_{{\cdot}\leftarrow i}^{H}$ & $\sum_{i=1}^{N}{\tilde{d}_{i1}^{gH}},i \neq 1$ & $\sum_{i=1}^{N}{\tilde{d}_{i2}^{gH}},i \neq 2$ & \multicolumn{2}{c}{$\cdots$} & $\sum_{i=1}^{N}{\tilde{d}_{iN}^{gH}},i\neq N$ & $1/N \sum_{i,j=1,j \neq i}^{N} {\tilde{d}_{ij}^{gH}}$\\
\bottomrule
\end{tabular}
\end{table}

\section{Overview of the connectedness}

By setting the predictive horizon  $H = 10$ and the lag order $p = 2$, we first estimate the volatility connectedness measures within 28 sectors in the entire period. The corresponding pairwise directional connectedness, from-connectedness, to-connectedness, net-connectedness, and total connectedness are listed in table \ref{Tb:con:static}. The pairwise directional connectedness measures between sectors form the volatility connectedness matrix. One can find that all the elements in connectedness matrix are in the range of $[1\%, 13\%]$. And each diagonal element in the matrix represents each sector's own connectedness ($C_{i \leftarrow i}$). One can see that in each row or column the diagonal volatility connectedness is the largest, indicating that the largest volatility shocks of each sector originate from itself. We also notice that the bank sector and non-bank financial sector exhibit the strongest self volatility connectedness (namely, 12.94\% and 11.72\%) and the chemical sector has the smallest self volatility connectedness (5.48\%), meaning that the self shocks in financial sectors are greater than those in other sectors. One can also find that self volatility connectedness ($C_{i \leftarrow i}$) is much less than its from-connectedness ($C_{\cdot \leftarrow i}$) for all sectors, suggesting that the volatility connectedness of each sector is dominated by the total external shocks from other sectors. The non-diagonal elements in the connectedness matrix represent the volatility shocks of sector $i$ coming from sector $j$. One can find that the volatility spillover from bank to commercial trade has the minimum value of 1.14\% and the volatility connectedness from electrical equipment to building materials achieves the highest value of 5.91\%. Furthermore, we also observe that there are pairs of sectors having relatively high spillover effects, such as the pair of bank and non-bank finance and of mechanical equipment and building materials. This can be explained by that bank and non-bank finance sectors form the financial industry and mechanical equipment and building materials have very strong connections to engineering constructions.

\begin{figure}[htpb]
\centering
\includegraphics[width=16cm]{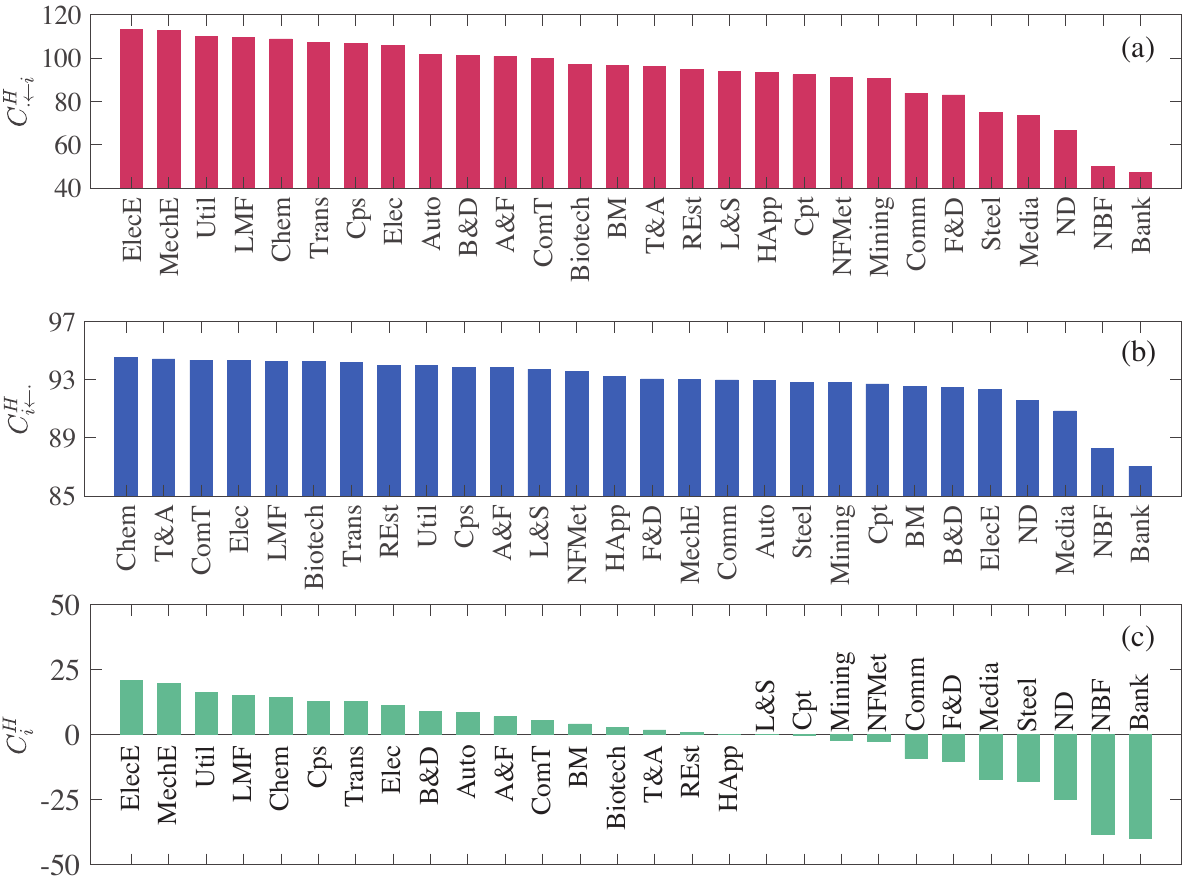}
\caption{\label{Fig:Evo:ConRank} Ranking plots of  (a) to-connectedness $C_{{\cdot}\leftarrow i}^{H}$, (b) from-connectedness $C_{i\leftarrow{\cdot}}^{H}$, and (c) net-connectedness $C_i^H$. }
\end{figure}

\begin{landscape}
\begin{table}[htp]
\setlength \tabcolsep{1.2pt}
\centering 
\scriptsize
\caption{\label{Tb:con:static} Volatility connectedness within 28 sectors in the Chinese stock markets during the sample period from 4 January 2000 to 31 December 2019. The results are obtained by setting the predictive horizon $H$ and the VAR lag order as 10 and 2 days. The sub-matrix from sector A{\&}F to sector MechE reports the pairwise directional volatility connectedness between sector $i$ and $j$ (Eq.~(\ref{Eq:Dgh:norm:row})), standing for the 10-day-ahead forecast error variance spilling from sector $j$ to sector $i$. ``From'', ``To'', and ``Net'' correspond to the from-connectedness (Eq.~(\ref{Eq:from:con})), to-connectedness (Eq.~(\ref{Eq:to:con})), and net-connectedness (Eq.~(\ref{Eq:net:con})) of sector $i$, respectively. The number in bold is the total connectedness (Eq.~(\ref{Eq:net:total})).}
\vskip 2.5mm
\begin{tabular}{lr@{.}lr@{.}lr@{.}lr@{.}lr@{.}lr@{.}lr@{.}lr@{.}lr@{.}lr@{.}lr@{.}lr@{.}lr@{.}lr@{.}lr@{.}lr@{.}lr@{.}lr@{.}lr@{.}lr@{.}lr@{.}lr@{.}lr@{.}lr@{.}lr@{.}lr@{.}lr@{.}lr@{.}lr@{.}l}
\toprule
industry  & \multicolumn{2}{c}{A\&F}  & \multicolumn{2}{c}{Mining}  &  \multicolumn{2}{c}{Chem}  
& \multicolumn{2}{c}{Steel}  & \multicolumn{2}{c}{NFMet}  & \multicolumn{2}{c}{Elec}  
& \multicolumn{2}{c}{HApp}  & \multicolumn{2}{c}{F\&D}  & \multicolumn{2}{c}{T\&A}  
& \multicolumn{2}{c}{LMF} & \multicolumn{2}{c}{Biotech} & \multicolumn{2}{c}{Util}  & \multicolumn{2}{c}{Trans}  
& \multicolumn{2}{c}{REst} & \multicolumn{2}{c}{ComT} & \multicolumn{2}{c}{L\&S} 
& \multicolumn{2}{c}{Cps}  & \multicolumn{2}{c}{BM} & \multicolumn{2}{c}{B\&D} 
& \multicolumn{2}{c}{ElecE} & \multicolumn{2}{c}{ND} & \multicolumn{2}{c}{Cpt}  & \multicolumn{2}{c}{Media}  
& \multicolumn{2}{c}{Comm} & \multicolumn{2}{c}{Bank} & \multicolumn{2}{c}{NBF} 
& \multicolumn{2}{c}{Auto}  & \multicolumn{2}{c}{MechE} & \multicolumn{2}{c}{\textbf{From}} \\
\midrule
 A\&F  & 6&18  & 3&22  & 4&52  & 2&78  & 3&64  & 4&36  & 3&65  & 3&43  & 4&27  & 4&37  & 4&46  & 4&27  & 4&17  & 3&81  & 4&37  & 4&04  & 4&15  & 3&19  & 3&27  & 3&75  & 2&39  & 3&06  & 2&36  & 2&92  & 1&27  & 1&40  & 3&06  & 3&65  & 93&82 \\
 Mining  & 3&45  & 7&20  & 4&40  & 3&59  & 4&29  & 3&81  & 3&67  & 3&47  & 3&57  & 4&00  & 3&68  & 4&63  & 4&39  & 3&93  & 3&99  & 3&43  & 3&86  & 3&11  & 3&42  & 3&67  & 2&11  & 2&77  & 2&12  & 2&85  & 1&69  & 1&80  & 3&36  & 3&75  & 92&80 \\
 Chem  & 4&22  & 3&59  & 5&48  & 3&25  & 3&75  & 4&28  & 3&57  & 3&52  & 4&48  & 4&56  & 4&47  & 4&69  & 4&46  & 3&90  & 4&56  & 3&98  & 4&08  & 3&05  & 3&16  & 3&58  & 2&16  & 2&85  & 2&16  & 2&81  & 1&32  & 1&47  & 3&15  & 3&46  & 94&52 \\
 Steel & 3&73  & 4&17  & 4&65  & 7&18  & 3&70  & 3&38  & 3&35  & 3&70  & 4&10  & 4&18  & 4&10  & 4&92  & 4&94  & 4&07  & 4&04  & 3&44  & 3&42  & 2&97  & 3&36  & 3&24  & 2&12  & 2&38  & 1&89  & 2&60  & 2&10  & 1&66  & 3&27  & 3&34  & 92&82 \\
 NFMet  & 3&97  & 4&16  & 4&47  & 3&06  & 6&41  & 4&08  & 3&51  & 3&20  & 3&81  & 4&18  & 3&79  & 4&09  & 4&25  & 3&96  & 4&02  & 3&67  & 4&02  & 3&37  & 3&42  & 3&90  & 2&50  & 3&01  & 2&22  & 2&74  & 1&32  & 1&66  & 3&35  & 3&86  & 93&59\\
 Elec  & 4&20  & 3&34  & 4&40  & 2&48  & 3&59  & 5&69  & 3&84  & 3&08  & 3&90  & 4&48  & 4&12  & 4&21  & 4&05  & 3&53  & 4&14  & 3&75  & 4&69  & 3&21  & 3&27  & 3&96  & 2&45  & 3&84  & 2&73  & 3&31  & 1&15  & 1&48  & 3&30  & 3&81  & 94&31\\
 HApp  & 3&81  & 3&59  & 4&11  & 2&77  & 3&48  & 4&25  & 6&80  & 3&45  & 3&51  & 4&29  & 3&85  & 4&37  & 4&35  & 3&59  & 3&76  & 3&45  & 4&35  & 3&15  & 3&42  & 3&83  & 2&06  & 3&26  & 2&55  & 3&08  & 1&93  & 1&77  & 3&54  & 3&63  & 93&20 \\
 F\&D  & 4&11  & 3&86  & 4&57  & 3&32  & 3&52  & 3&90  & 3&84  & 6&95  & 4&20  & 4&17  & 4&77  & 4&55  & 4&37  & 3&89  & 4&65  & 3&97  & 3&52  & 2&73  & 3&07  & 3&28  & 1&95  & 2&55  & 2&05  & 2&64  & 1&59  & 1&92  & 2&87  & 3&20  & 93&05 \\
 T\&A  & 4&43  & 3&28  & 4&94  & 3&10  & 3&52  & 4&22  & 3&40  & 3&53  & 5&57  & 4&56  & 4&71  & 4&56  & 4&24  & 3&93  & 4&74  & 4&13  & 4&00  & 3&13  & 3&19  & 3&58  & 2&18  & 2&89  & 2&32  & 2&64  & 1&21  & 1&33  & 3&11  & 3&55  & 94&43 \\
 LMF  & 4&07  & 3&31  & 4&56  & 2&94  & 3&54  & 4&36  & 3&67  & 3&21  & 4&14  & 5&72  & 4&04  & 4&31  & 4&23  & 3&68  & 4&18  & 3&85  & 4&82  & 3&25  & 3&27  & 3&97  & 2&14  & 3&31  & 2&44  & 2&97  & 1&43  & 1&36  & 3&41  & 3&83  & 94&28\\
 Biotech  & 4&57  & 3&38  & 4&92  & 3&10  & 3&52  & 4&41  & 3&72  & 4&01  & 4&68  & 4&46  & 5&76  & 4&66  & 4&36  & 3&89  & 4&81  & 4&19  & 3&93  & 2&80  & 2&95  & 3&42  & 2&04  & 2&83  & 2&24  & 2&69  & 1&15  & 1&28  & 2&93  & 3&28  & 94&24 \\
 Util  & 3&95  & 3&94  & 4&70  & 3&45  & 3&39  & 4&04  & 3&83  & 3&61  & 4&08  & 4&23  & 4&28  & 6&00  & 4&82  & 3&68  & 4&41  & 3&77  & 4&16  & 2&92  & 3&37  & 3&55  & 2&17  & 2&86  & 2&24  & 2&84  & 1&56  & 1&50  & 3&18  & 3&47  & 94&00 \\
 Trans  & 3&89  & 3&82  & 4&51  & 3&53  & 3&59  & 4&02  & 3&90  & 3&54  & 3&85  & 4&34  & 4&00  & 4&95  & 5&82  & 3&89  & 4&29  & 3&78  & 4&11  & 2&99  & 3&52  & 3&52  & 2&12  & 2&83  & 2&20  & 2&84  & 1&82  & 1&62  & 3&22  & 3&50  & 94&18 \\
 REst  & 4&08  & 3&72  & 4&46  & 3&22  & 3&81  & 3&93  & 3&68  & 3&48  & 4&07  & 4&25  & 4&02  & 4&24  & 4&38  & 6&00  & 4&25  & 4&06  & 3&97  & 3&25  & 3&28  & 3&50  & 2&14  & 2&74  & 2&16  & 2&78  & 1&89  & 1&80  & 3&22  & 3&61  & 94&00 \\
 ComT  & 4&39  & 3&76  & 4&93  & 3&02  & 3&68  & 4&38  & 3&67  & 3&90  & 4&57  & 4&45  & 4&76  & 4&79  & 4&57  & 3&99  & 5&66  & 4&15  & 4&12  & 2&70  & 2&92  & 3&34  & 2&05  & 2&84  & 2&07  & 2&66  & 1&14  & 1&41  & 2&77  & 3&28  & 94&34 \\
 L\&S  & 4&32  & 3&36  & 4&54  & 2&77  & 3&53  & 4&15  & 3&52  & 3&53  & 4&20  & 4&47  & 4&30  & 4&33  & 4&21  & 4&08  & 4&36  & 6&28  & 4&10  & 3&09  & 3&21  & 3&81  & 2&13  & 2&95  & 2&54  & 2&79  & 1&29  & 1&41  & 3&13  & 3&58  & 93&72\\
 Cps  & 3&93  & 3&28  & 4&12  & 2&47  & 3&44  & 4&57  & 3&80  & 2&82  & 3&62  & 4&90  & 3&61  & 4&26  & 4&05  & 3&49  & 3&84  & 3&66  & 6&17  & 3&53  & 3&51  & 4&23  & 2&30  & 3&69  & 2&71  & 3&16  & 1&50  & 1&62  & 3&64  & 4&06  & 93&83 \\
 BM  & 3&40  & 2&94  & 3&32  & 2&28  & 3&09  & 3&47  & 3&08  & 2&38  & 2&98  & 3&77  & 2&77  & 3&34  & 3&27  & 3&15  & 2&83  & 3&03  & 3&88  & 7&45  & 5&42  & 5&91  & 3&04  & 3&96  & 3&26  & 3&40  & 1&99  & 1&98  & 4&97  & 5&67  & 92&55 \\
 B\&D  & 3&39  & 3&24  & 3&37  & 2&52  & 3&03  & 3&52  & 3&24  & 2&65  & 2&92  & 3&66  & 2&86  & 3&84  & 3&87  & 3&12  & 3&07  & 3&02  & 3&92  & 5&02  & 7&51  & 5&47  & 2&89  & 3&83  & 2&98  & 3&33  & 2&13  & 2&18  & 4&30  & 5&13  & 92&49 \\
 ElecE  & 3&49  & 2&82  & 3&43  & 2&19  & 3&14  & 3&70  & 3&16  & 2&40  & 2&97  & 3&90  & 2&94  & 3&54  & 3&27  & 2&93  & 2&98  & 3&21  & 3&98  & 5&05  & 5&05  & 7&65  & 3&16  & 4&39  & 3&43  & 3&36  & 1&72  & 1&91  & 4&73  & 5&51  & 92&35 \\
 ND  & 3&54  & 2&74  & 3&38  & 2&26  & 3&23  & 3&74  & 2&74  & 2&38  & 3&00  & 3&28  & 2&89  & 3&46  & 3&32  & 2&95  & 3&00  & 3&02  & 3&45  & 4&54  & 4&73  & 5&40  & 8&42  & 3&98  & 3&18  & 3&44  & 1&48  & 2&50  & 4&53  & 5&40  & 91&58\\
 Cpt  & 3&43  & 2&79  & 3&36  & 2&04  & 3&09  & 4&46  & 3&26  & 2&32  & 2&96  & 3&89  & 3&01  & 3&49  & 3&14  & 2&82  & 3&10  & 3&05  & 4&26  & 3&99  & 4&07  & 5&23  & 2&96  & 7&29  & 4&70  & 4&17  & 1&51  & 1&71  & 4&71  & 5&18  & 92&71 \\
 Media  & 3&27  & 2&62  & 3&09  & 1&93  & 2&71  & 4&00  & 2&95  & 2&18  & 2&93  & 3&58  & 2&82  & 3&31  & 3&08  & 2&66  & 2&83  & 3&12  & 3&88  & 4&06  & 3&95  & 5&02  & 2&87  & 5&78  & 9&16  & 4&07  & 1&67  & 1&95  & 4&93  & 5&57  & 90&84 \\
 Comm  & 3&59  & 3&15  & 3&57  & 2&39  & 3&07  & 4&13  & 3&39  & 2&64  & 3&01  & 3&86  & 3&08  & 3&73  & 3&56  & 3&12  & 3&18  & 3&20  & 4&01  & 3&81  & 3&92  & 4&50  & 2&85  & 4&67  & 3&63  & 7&02  & 1&91  & 1&80  & 4&37  & 4&82  & 92&98 \\
 Bank  & 2&47  & 3&19  & 2&82  & 3&09  & 2&52  & 2&42  & 3&53  & 2&75  & 2&23  & 3&21  & 2&06  & 3&34  & 3&64  & 3&56  & 2&18  & 2&30  & 3&18  & 4&01  & 4&62  & 4&12  & 2&24  & 3&06  & 2&79  & 3&39  & 12&94  & 5&14  & 4&91  & 4&26  & 87&06 \\
 NBF  & 2&53  & 3&26  & 2&94  & 2&38  & 2&96  & 2&91  & 3&18  & 2&83  & 2&32  & 2&79  & 2&19  & 3&21  & 3&21  & 3&21  & 2&53  & 2&55  & 3&18  & 3&93  & 4&63  & 4&56  & 3&36  & 3&21  & 3&13  & 3&01  & 5&19  & 11&72  & 4&43  & 4&66  & 88&28 \\
 Auto  & 3&17  & 2&98  & 3&37  & 2&51  & 3&01  & 3&50  & 3&27  & 2&42  & 2&96  & 3&82  & 2&83  & 3&54  & 3&46  & 3&05  & 2&82  & 2&93  & 3&84  & 4&81  & 4&48  & 5&39  & 3&07  & 4&36  & 3&72  & 3&61  & 2&35  & 2&11  & 7&08  & 5&55  & 92&92 \\
 MechE  & 3&36  & 3&04  & 3&31  & 2&26  & 3&13  & 3&65  & 3&01  & 2&47  & 2&96  & 3&76  & 2&79  & 3&51  & 3&37  & 3&05  & 3&00  & 2&99  & 3&86  & 5&02  & 4&87  & 5&57  & 3&26  & 4&42  & 3&86  & 3&64  & 1&81  & 2&03  & 5&05  & 6&96  & 93&04\\
 \textbf{To} & 100&76  & 90&55  & 108&79  & 74&70  & 90&96  & 105&64  & 93&44  & 82&93  & 96&28  & 109&42  & 97&19  & 110&12  & 107&04  & 94&90  & 99&94  & 93&71  & 106&74  & 96&69  & 101&35  & 113&28  & 66&69  & 92&35  & 73&67  & 83&76  & 47&14  & 49&82  & 101&43  & 112&63 &  \textbf{92}& \textbf{93} \\
 \textbf{Net} & 6&94  & -2&25  & 14&27  & -18&12  & -2&63  & 11&33  & 0&24  & -10&13  & 1&86  & 15&14  & 2&95  & 16&13  & 12&85  & 0&90  & 5&60  & -0&01  & 12&92  & 4&14  & 8&86  & 20&93  & -24&89  & -0&36  & -17&17  & -9&22  & -39&92  & -38&47  & 8&52  & 19&60 & \multicolumn{2}{c}{} \\
\bottomrule
\end{tabular}
\begin{tablenotes}
\item{Note:} Agriculture and forestry (A{\&}F), Mining (Mining), Chemical (Chem), Steel (Steel), Non-ferrous metals (NFMet), Electronic (Elec), Household appliances (HApp), Food and drink (F\&D), Textile and apparel (T\&A), Light manufacturing (LMF), Biotechnology (Biotech), Utilities (Util), Transportation (Trans), Real estate (REst), Commercial trade (ComT), Leisure and services (L\&S), Composite (Cps), Building materials (BN), Building and decoration (B\&D), Electrical equipment (ElecE), National defense (ND), Computer (Cpt), Media (Media), Communications (Comm), Bank (Bank), Non-bank financial (NBF), Automobile (Auto), and Mechanical equipment (MechE).
\end{tablenotes}
\end{table}
\end{landscape}

Table~\ref{Tb:con:static} reports the to-connectedness, from-connectedness, and net-connectedness for each sector. One can find that the to-connectedness ranges from 47.14\% to 113.28\%, the maximum and minimum values of the from-connectedness are 94.52\% and 87.06\%, and the net-connectedness spans from -39.92\% to 20.93\%. For better visibility of the rank of the three connectedness measures, we plot the from-connectedness, to-connectedness, and net-connectedness in descending order in Fig.~\ref{Fig:Evo:ConRank}. For to-connectedness, the sectors like mechanical equipment, electrical equipment, and utilities have the strongest volatility spillover effects on the other sectors and their spillover weights are as high as 110\%, as evidenced in Fig.~\ref{Fig:Evo:ConRank} (a). This can be explained by that these sectors are generally basic industries, accounting for a great part of the economic structure in China and having strong relevance with the other sectors. However, the sectors like bank and non-bank finance have the lowest to-connectedness with values of 47.14\% and 49.82\%, indicating that the financial sectors have the minimum risk spillover effects on the other sectors. As shown in  Fig.~\ref{Fig:Evo:ConRank} (b), the from-connectedness varies in a very narrow range from 87.06\% to 94.52\%, suggesting that for each sector around 90\% of its volatility shocks are contributed by other sectors. One can see that the sectors of bank and non-bank finance relatively have the least from-connectedness with values of  87.06\% and 88.28\%, however, the gaps to the maximum from-connectedness are small, suggesting that the financial sectors receive comparable volatility shocks comparing with the other sectors. We further estimate the net-connectedness by calculating the difference between to-connectedness and from-connectedness. The positive (respectively, negative) net-connectedness means that a sector is a risk ``transmitter'' (``receiver''), transmitting its shocks to (accepting shocks from) other sectors. As shown in Fig.~\ref{Fig:Evo:ConRank} (c), one can find that 17 sectors, including mechanical equipment, electrical equipment, utilities, and so on, are risk transmitters and 11 sectors, such as national defence, non-bank finance, bank, and to list a few, are risk receivers.

Finally, we estimate the total connectedness to uncover the average level of risk spillover within sectors. One can see that the total risk spillover score reaches 92.93\%, indicating that sectors in the Chinese stock markets exhibit extremely high intensity of spillovers, comparing with the total connectedness of 33.46\%, 45.79\%, and 76.93\% within sectors in the Eurozone credit markets \citep{Shahzad-Bouri-Arreola-Roubaud-Bekiros-2019-AE}, in the US CDS markets \citep{Fonseca-Ignatieva-2018-AE} and the US credit markets \citep{Collet-Ielpo-2018-JBF}, respectively.

\section{Evolution of the total connectedness}

\begin{figure}[t]
\centering
\includegraphics[width=16cm]{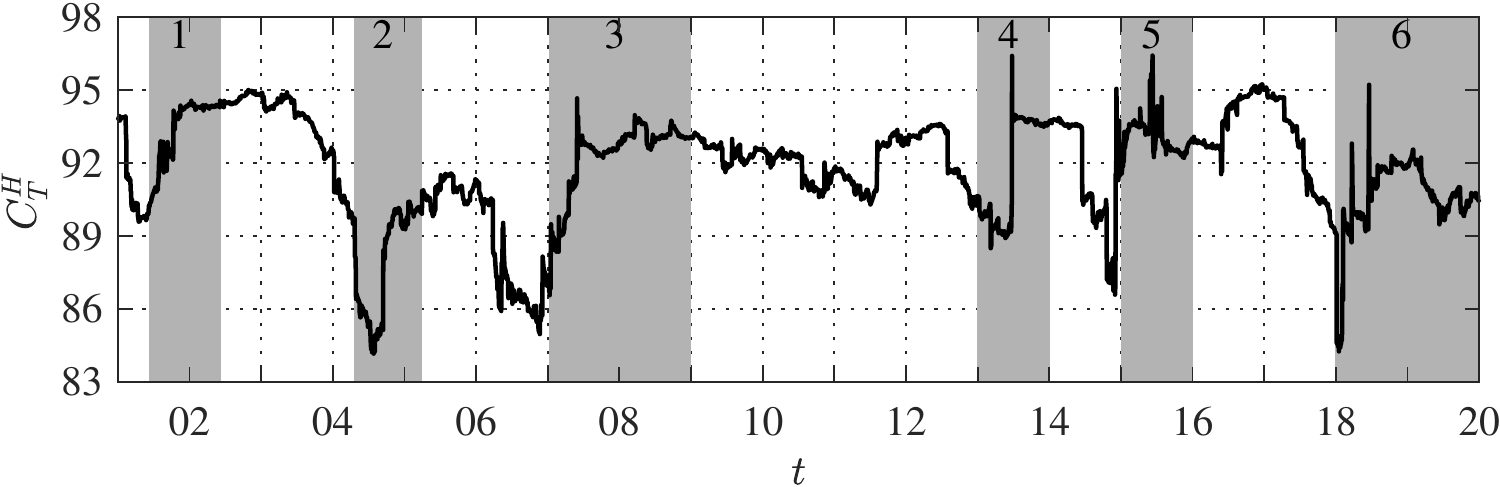} 
\caption{\label{Fig:Evo:totCon} Plots of the evolution of the total connectedness within sectors in the period from January 2, 2001 to December 31, 2019. The total connectedness is obtained from the rolling window analysis with a window size of 240 days, a moving step of one days, a lag order of 2, and a predictive horizon of 10 days. The six shadow areas highlight the peaks associated with the events that have great impacts on the Chinese stock markets.}
\end{figure}

In this section, a rolling window analysis is performed to uncover the evolution of the risk spillovers within sectors in the Chinese stock markets. The window with a fixed size of 240 days rolls from January 4, 2001 to December 31, 2019 with a step of one day. In each window, the total connectedness is estimated with the same model parameters (the lag order $p = 2$ and predictive horizon $H = 10$) as the whole sample analysis. Fig.~\ref{Fig:Evo:totCon} illustrates the plot of the total connectedness. One can see that the total connectedness presents a volatile style with fluctuations between 84.43\% and 96.43\%. We also highlight the peaks associated with the events that have great impacts on the Chinese stock markets in shadow areas.

As shown in Fig.~\ref{Fig:Evo:totCon}, the first shadow locating in the period from mid 2001 to mid 2002 exceeds the score of 94\%, which can be ascribed to that the policy of reducing the state-owned shares in 2001 makes the Chinese stock markets shift into a bear state in the following two years. And at the beginning of 2004, the State Council introduced a policy to promote the reform and opening of the Chinese capital markets and further to encourage the development of the markets. However, such policy initiated a stock market decline lasting a period from April 2004 to June 2005 (Shadow 2), which made the total connectedness increase sharply during that period. On April 29, 2005, the reform of the shareholder structure in listed companies was launched, which stopped the falling of the stock markets, triggered the price increment in the following, and finally drove the market into a bubble state, agreeing well with the ``V'' shape of the total connectedness between mid-2005 and mid-2007. The total connectedness reached 92\% near the end of 2007, in consistent with the fact that the Chinese stock markets crashes in October of 2007 \citep{Jiang-Zhou-Sornette-Woodard-Bastiaensen-Cauwels-2010-JEBO}. Because of the 2008 global fianncial crisis (Shadow 3), the total connectedness also stayed in a state of high value in the following years.

Due to the high frequency of financial crisis (the subprime crisis 2007–2009, the global 2008 financial crisis, and the European sovereign debt crisis 2011–2015 \citep{Jiang-Wang-Canabarro-Podobnik-Xie-Stanely-Zhou-2018-QF}) around the world in recent years, the total connectedness fluctuated greatly in the Chinese stock markets. The total connectedness achieved the score of 96\% in June 2013 (Shadow 4), which was caused by the most serious ``money shortage'' (interbank liquidity crisis) in the financial industry. The overnight interest rate of the Shanghai Interbank Offered Rate (``Shibor'') rose to 13.44\%, the highest record since 2006. On June 24, 2013, the Chinese market index plunged 5.3\%, leading to a sharp peak in the total connectedness curve around that day. The money shortage was ended by the policy that the central bank introduced 416 billion Yuan into the market through SLF (Standard for Loan Facility). In 2015 (Shadow 5), the total connectedness significantly increased again and reached the score of 95\%. The Chinese stock markets ushered in a plunge in June, 2015, and more than thousands of stocks depreciated than 50\%. Many listed companies had suspended the share trading for different reasons to stop price dropping. In June 2018 (Shadow 6), the total connectedness exceeded the score of 95\% once again. On June 19, thousands of stocks slumped by the maximum 10\% daily limits, suggesting very high overall risk in the markets. There are many reasons which cause this market crash. One possible explanation could be the China-US trade conflicts in June 2018. Trump announced that it would impose a 10\% punitive tariff on 200-billion dollars of Chinese goods, which had a heavy impact on the Chinese stock markets. Another reason could be that the confidence of investors was greatly hurt by the seasoned equity offering of list companies.

\section{Evolution of the sector connectedness}

\begin{figure}[htpb]
\centering
\includegraphics[width=15cm]{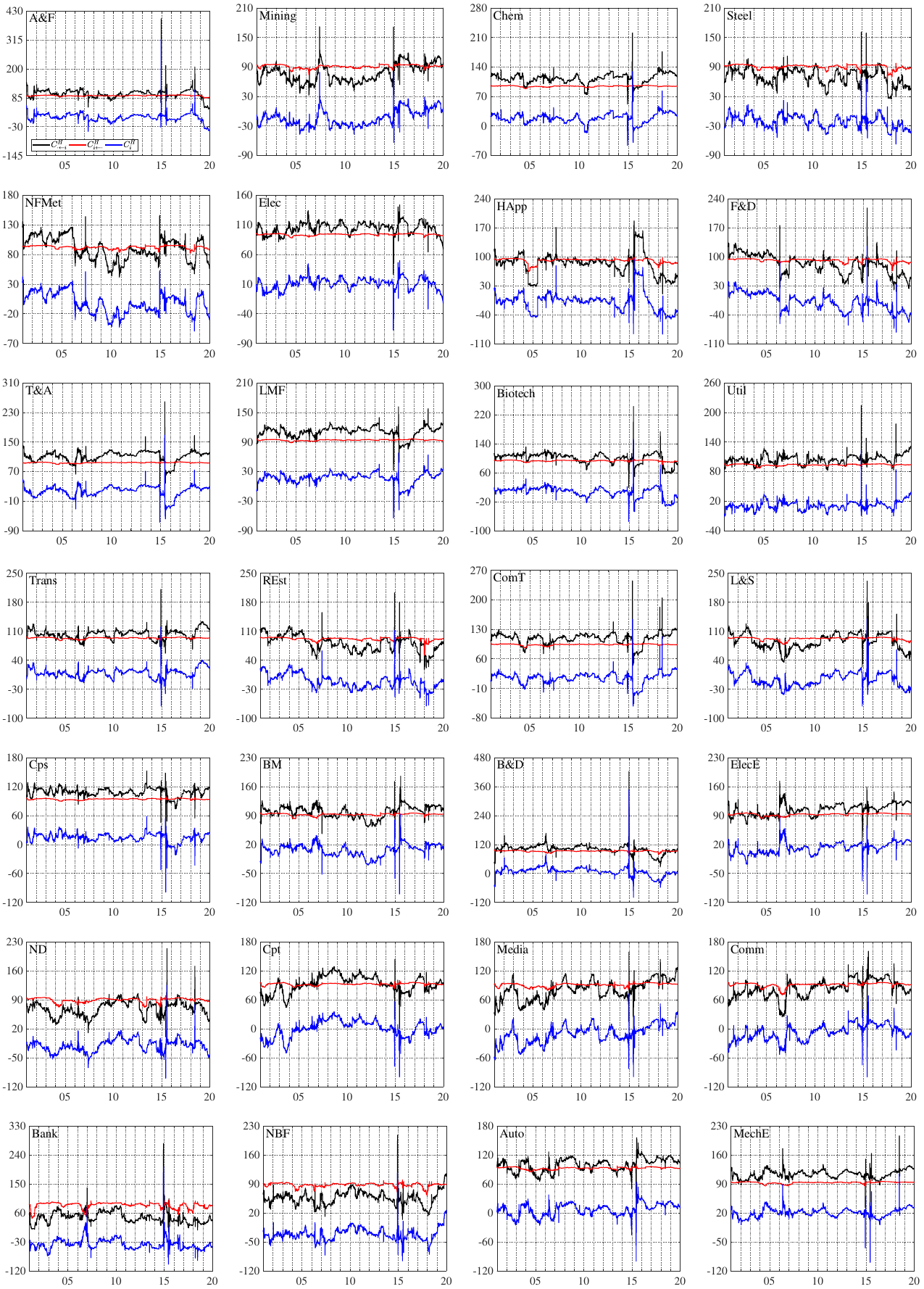}
\caption{\label{Fig:Evo:secCon} Evolution of from-connectedness, to-connectedness, and net-connectedness for 28 sectors from January 2, 2001 to December 31, 2019. The from-connectedness, to-connectedness, and net-connectedness are obtained from the rolling window analysis with a window size of 240 days, a moving step of one day, a lag order of 2, and a predictive horizon of 10 days. Each panel represents the results of one sector.}
\end{figure}

We further consider the evolution of to-connectedness, from-connectedness, net-connectedness for each sector.  Fig.~\ref{Fig:Evo:secCon} illustrates the evolution of three connectedness measures, which are also obtained by the rolling window analysis with a window size of 240 days, a moving step of one day, a lag order of 2, and a predictive horizon of 10 days. One can find that except the sectors of bank and non-bank finance, the from-connectedness of other sectors stays above the score of 80\% and varies in a very narrow range, as evidenced by the small gap between their maximum and minimum value. And the from-connectedness of bank and non-bank financial sectors exhibits volatile fluctuations and goes under the score of 80\% in some periods. This means that the financial industry is more sensitive and vulnerable to the risk events including policy changes, international economic situations, financial crash events, and to list a few. It is observed that to-connectedness is greatly larger than from-connectedness for all sectors, which indicates that no sector is solo in the markets and each sector has risk spillover to other sectors. One can also see that all sectors have peaks around the same time point, which corresponds to the events of market crashes. Such events are signals of risk release, thus increasing the to-connectedness of all sectors.

The net-connectedness is allowed to have both positive and negative values according to its definition. The positive and negative values represent the shocks transmitted to and received from other sectors, respectively. As shown in Fig.~\ref{Fig:Evo:secCon}, the net-connectedness and to-connectedness curves have peaks at the same time for all sectors, however the net-connectedness exhibits much greater fluctuations. For the sectors of agriculture \& forestry, chemical, electronic, textile and apparel, light manufacturing, biotechnology, utilities, transportation, commercial trade, composite, building materials, building \& decoration, electrical equipment, and mechanical equipment, their net-connectedness measures keep positive for most of the time, indicating that these sectors are risk transmitters. And the sectors of mining, steel, household appliances, real estate, national defence, media, communications, bank, and non-bank finance have negative net-connectedness for most of the time, implying that these sectors are risk receivers. The result that the real estate sector acts as a risk taker contrasts with our intention that the real estate should transmit risks to other sectors as it is a high risk industry. This can be explained by that as a pillar industry of China the real estate industry plays a critical role in the social development and urbanization process and in the industry chain of real estate, other sectors provide products and services to the real estate sector. As shown in Fig.~\ref{Fig:Evo:secCon}, we can find around 2015 bank and non-bank financial sectors change from risk receivers to risk transmitters, at the same time the sectors including electrical equipment, national defence, computer, and media act as risk receivers. It is observed that at the end of 2014 the net connectedness value of the bank sector approaches to 200\%  and the net-connectedness of the non-bank finance exceeds the score of 100\%. This means that the financial sector exhibits very strong external spillovers in that period, which cannot be uncovered in the entire sample analysis. Besides the bank and non-bank finance, the following sectors including agriculture \& forestry, chemical, food \& drink, textile and apparel, biotechnology, commercial trade, and leisure \& services are also risk transmitters. In June 2018, the US government announced the tariff list for the Chinese 50 billion imported goods, including the products from the sectors of semiconductors and chips, robotics and machinery, navigation and automation, and information and communication technology. As shown in Fig.~\ref{Fig:Evo:secCon}, one can see that the taxation incident has had a great impact on the related sectors. The sectors of communication and computer have changed from risk receivers to risk transmitters since June 2018. 

\begin{figure}[htpb]
\centering
\includegraphics[width=16cm]{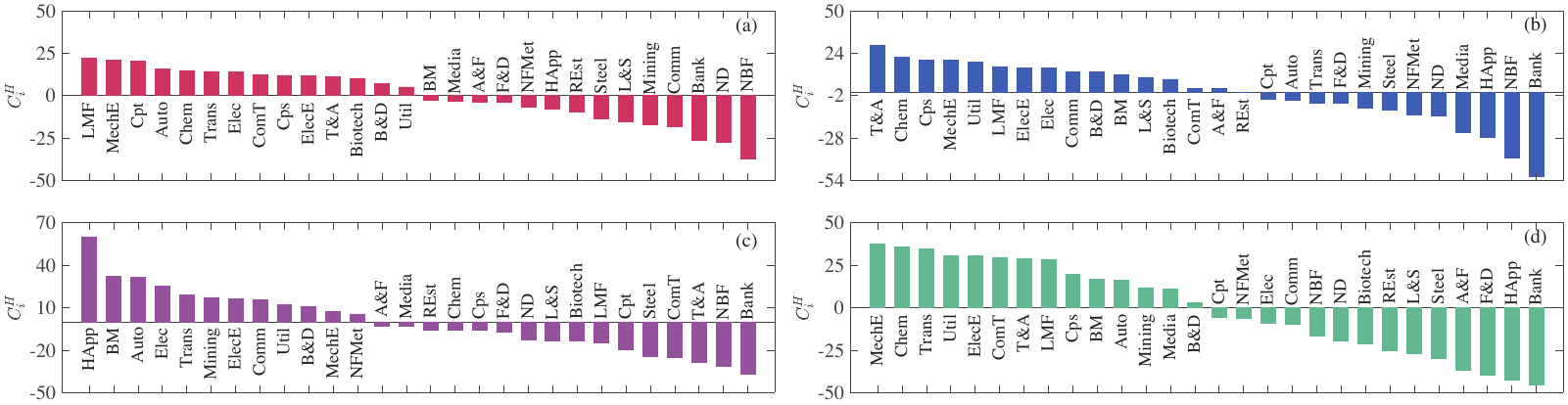}
\caption{\label{Fig:NetConRk:Risk:Events} Bar plots of the net-connectedness in descending order for the four subperiods. (a) The global financial crisis from 2007 to 2008. (b) The Chinese interbank liquidity crisis (money shortage) in 2013. (c) The Chinese stock market plunge in 2015. (d) The China-US trade war since 2018.}
\end{figure}

\section{Connectedness networks around extreme risk events}

In this section, we further perform the analysis of risk spillover within sectors in four subperiods corresponding to the shadow area 3, 4, 5, and 6 in Fig.~\ref{Fig:Evo:totCon}. The four subperiods are chosen because they cover the four typical extreme risk events in the Chinese stock markets, such as the global financial crisis from 2007 to 2008, the Chinese interbank liquidity crisis (money shortage) in 2013, the Chinese stock market plunge in 2015, and the China-US trade war since 2018. In each subperiod, we estimate the volatility connectedness matrix $\tilde{D}^{gH}$ within 28 sectors using the VAR model with the same model parameters as the full sample analysis and the rolling window analysis, which allow us to estimate the to-connectedness, from-connectedness, and net-connectedness for each subperiod. We thus plot the net-connectedness in descending order in Fig.~\ref{Fig:NetConRk:Risk:Events} for the four subperiods. In comparison of Fig.~\ref{Fig:Evo:ConRank} (c), one can see that in the four subperiods and entire period the sectors of bank and non-bank finance are always the top 3 risk receivers, indicating that the financial sectors are crucial for stabilizing the whole economic system. Moreover, the sectors of national defence, steel, and food and drink also act as the same role as the financial sectors. However, the top risk transmitters are not always the same and are dependent on the specific risk events. Specifically, the top 3 risk contributors are the sectors of light manufacturing, mechanical equipment, and computer (respectively, the sectors of textile and apparel, chemical, and composite, the sectors of household appliances, building and materials, and automobile, and the sectors of mechanical equipment, chemistry, and transportation) during the subperiods of the global financial crisis (the Chinese interbank liquidity crisis, the Chinese stock market plunge, and the China-US trade war). 

To illustrate the risk transmitting path, we take the latest risk event, the China-US trade war, as an example . On March 22, 2018, the US government announced to add 25\% tariff on the aluminium, iron and steel imported from China. In June 2018, the US government put more Chinese goods onto the tariff list, including semiconductors and chips, robotics and machinery, navigation and automation, and information and communication technology. Thus, the related sectors which produce the products on the tariff list were severely impacted. For example, adding tariffs on the machinery will directly increase the risk of the mechanical equipment sector. As a role of basic industry, the chemical sector offers raw materials to the sectors receiving the shocks from the trade war, and its risk gradually increased. Furthermore, the trade war certainly reduces the orders, thus adds the risk exposure of the transportation sector. The accumulating of the risks in other sectors are finally accepted by the sectors of bank, household appliances, food and drink, and other sectors.

\begin{figure}[htpb]
\centering
\includegraphics[width=16cm]{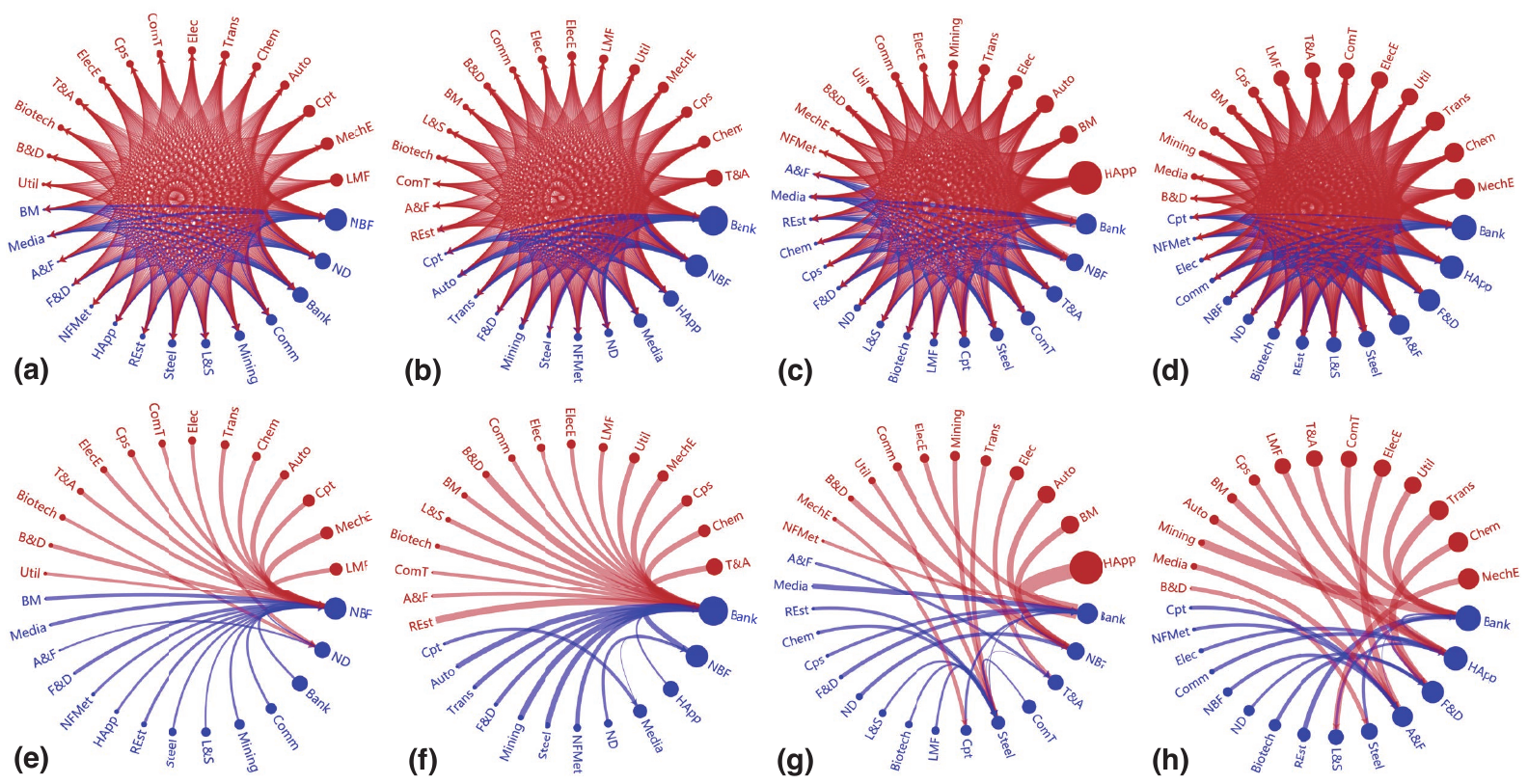}
\caption{\label{Fig:Net:Risk:Events} Plots of the connectedness networks associated with four typical extreme risk events occurred in four subperiods. The node size is proportional to the value of net-connectedness. The nodes in red and blue represent risk contributors and risk receivers, respectively. The link width is proportional to the value of the pairwise directional connectedness. The links in red and blue are sourced from risk contributors and risk receivers, respectively. (a) The global financial crisis from 2007 to 2008. (b) The Chinese interbank liquidity crisis (money shortage) in 2013. (c) The Chinese stock market plunge in 2015. (d) The China-US trade war since 2018. (e), (f), (g), and (h) are subgraphs of (a), (b), (c), and (d), respectively. In each subgraph, only the outgoing link with maximum net pairwise connectedness is shown for each node.}
\end{figure}

\begin{figure}[htpb]
\centering
\includegraphics[width=16cm]{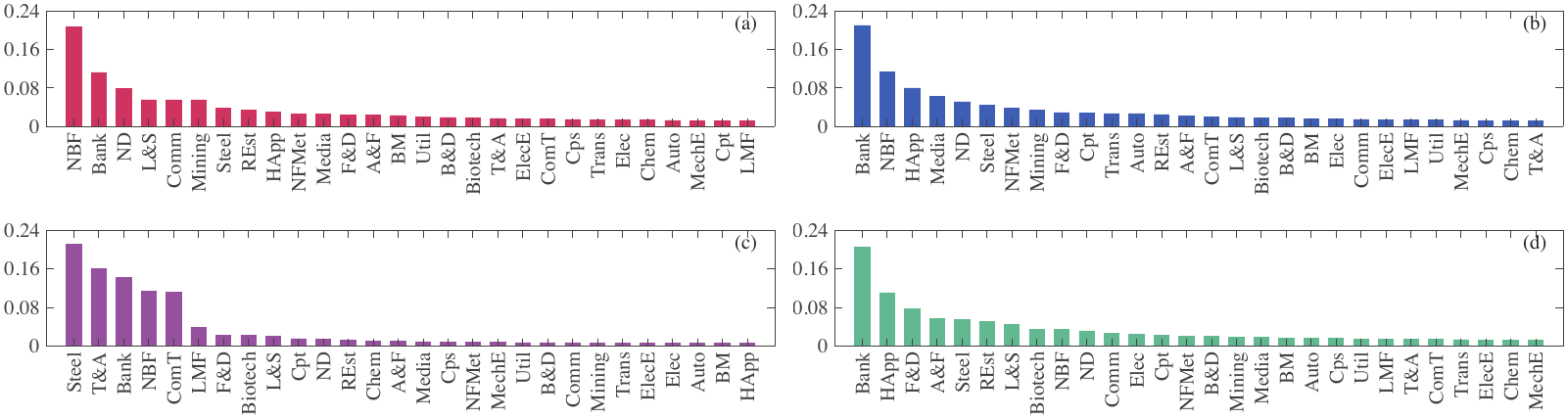}
\caption{\label{Fig:PgRk:Risk:Events} Bar plots of the PageRank score in descending order for the net pairwise connectedness networks obtained from the four subperiods. (a) The global financial crisis from 2007 to 2008. (b) The Chinese interbank liquidity crisis (money shortage) in 2013. (c) The Chinese stock market plunge in 2015. (d) The China-US trade war since 2018.}
\end{figure}

Based on the connectedness matrix $\tilde{D}^{gH}$, a net pairwise connectedness matrix is $\tilde{C}^{gH}$ defined as,
\begin{equation}
\tilde{c}_{ij}^{gH}= 
\begin{cases} 
\tilde{d}_{ij}^{gH} - \tilde{d}_{ji}^{gH}, \quad  & \tilde{d}_{ij}^{gH} - \tilde{d}_{ji}^{gH} > 0, \\ 
  0, \quad  &  \tilde{d}_{ij}^{gH} - \tilde{d}_{ji}^{gH} \le 0.
\end{cases} 
\label{Eq:Net:Pair:Con:Mat}
\end{equation}
Figs.~\ref{Fig:Net:Risk:Events} (a - d) illustrate the net connectedness networks visualized from the pairwise directional connectedness matrix for the four subperiods. The node size and link width are proportional to the net-connectedness and the net pairwise directional connectedness, respectively. The nodes in red and blue represent the risk contributors and the risk receivers. Figs.~\ref{Fig:Net:Risk:Events} (e - h) illustrate the subgraphs of the net connectedness networks in the four subperiods. The subgraphs are obtained through visualizing the outgoing links, representing the risk transmitted from the source to the target, with the maximum connectedness for each node in each subperiod. One can see that for the global financial crisis in 2008 (respectively, the Chinese interbank liquidity crisis in 2013, the Chinese stock market plunge in 2015, the China-US trade war since 2018) the important risk absorbers are the sectors of non bank finance and national defence (the sectors of banks and non bank finance, the sectors of bank and household appliances), which receive the maximum net risk spillover from the other sectors. To assess the importance of sectors, we further apply the algorithm of PageRank to estimate the centrality score of the sectors in each net connectedness network. Fig.~\ref{Fig:PgRk:Risk:Events} illustrates the PageRank score of the sectors in descending order for the four subperiod networks. Again, one can see that the sectors of bank and non bank finance have the top PageRank score, indicating that the financial sectors act as buffer roles in turbulent periods and keeping them stabilized and functional is critical in managing systematic risk.

\section{Robust tests}

\begin{figure}[htp]
\centering
\includegraphics[width=16cm]{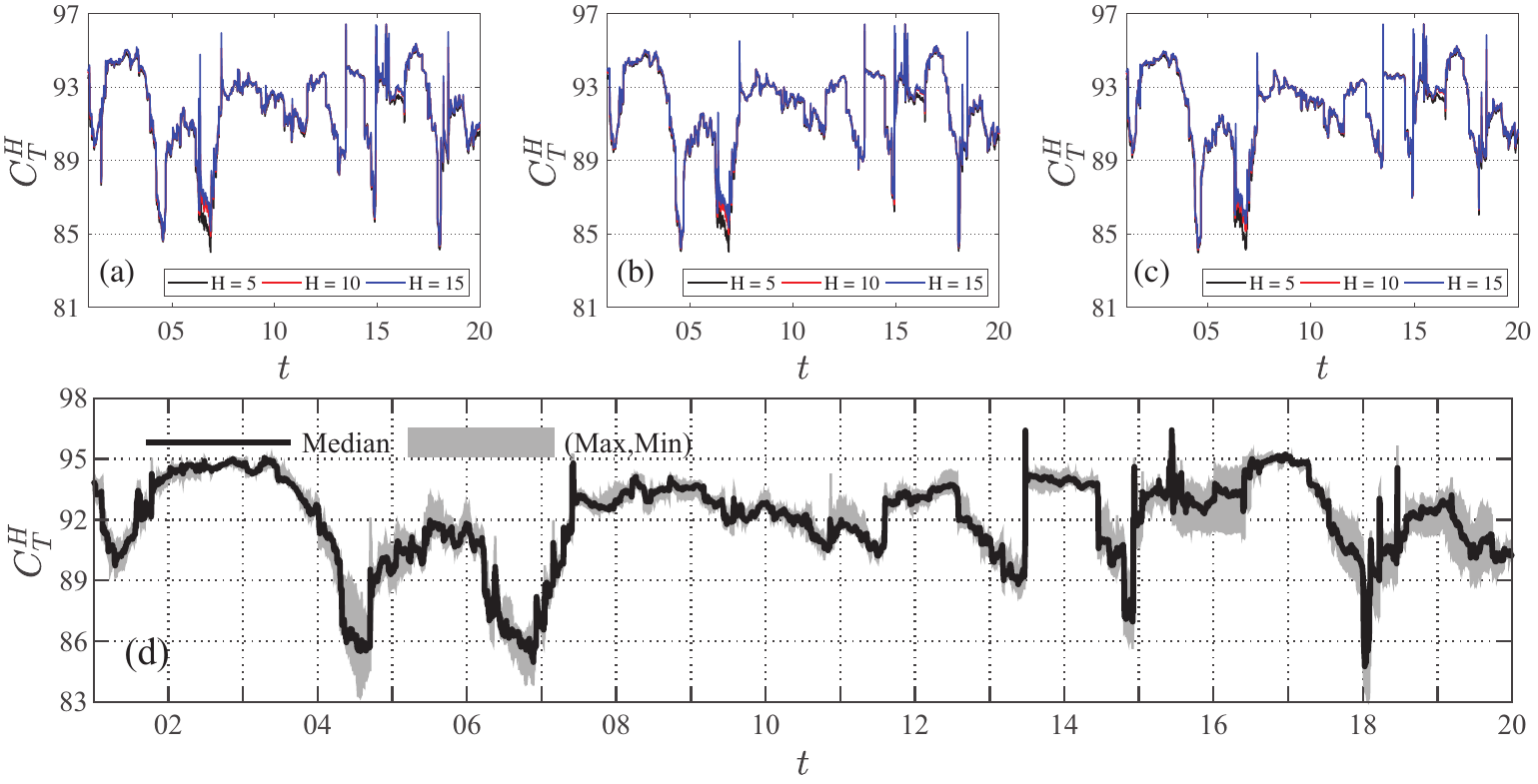} 
\caption{\label{Fig:totCon:rubtests} Results of rubost tests. (a) Plots of the tot-connectedness for $p = 2$, $W = 220$, and different $H$. (b) Plots of the tot-connectedness for $p = 2$, $W = 240$, and different $H$. (c) Plots of the tot-connectedness for $p = 2$, $W = 260$, and different $H$. (d) Plots of the tot-connectedness for $W = 220$, $H = 10$, and different $p$. The bounds of shadow area correspond to the maximum and minimum value of the tot-connectedness for different lags. }
\end{figure}

Our connectedness measure is constructed based on the generalized variance decomposition framework of VAR($p$) model, which has three parameters including the lag order $p$, the predictive horizon $H$, and the size of rolling window $W$. 
In above analysis, the model parameters are set as $p = 2$, $H = 10$, and $W = 240$, respectively. To perform the robust tests, one can change the values of the model parameters to check their influences on the results. By fixing $p = 2$, the rolling window size $W$ is set as 220 days, 240 days, and 260 days, and the predictive horizon $H$ is set as 5 days, 10 days, and 15 days. The results are shown in Fig.~\ref{Fig:totCon:rubtests}. One can find that the tot-connectedness curves in the three panels are almost overlapping on the same curve, indicating that our results are independent on the model parameters $W$ and $H$. We then redo the analysis by setting the lag order $p$ as 1, 2, 3, 4, 5 and fixing $W = 240$ and $H = 10$. The results are shown in Fig.~\ref{Fig:totCon:rubtests} (d). The upper and lower borders of the shadow area correspond to the maximum and minimum value of the tot-connectedness for different lags. We also plot the median value of tot-connectedness in black solid line. It is observed that the results are not sensitive to the changes of lags. In summary, the connectedness measure is robust to the changes of model parameters.

\section{Conclusion}

Based on the generalized variance decomposition framework of VAR model, we analyze the volatility connectedness within sectors in the Chinese stock markets. First, we construct a connectedness matrix based on the full data sample, which gives the from-connectedness, to-connectedness, net-connectedness among sectors, as well as tot-connectedness. It is found that the greatest risk contributor to each sector is itself as the largest element in each row of the connectedness matrix all locates on the diagonal line. 17 Sectors (mechanical equipment, electrical equipment, utilities, and so on) are the risk transmitters and 11 sectors (national defence, bank, and non-bank finance and to list a few) are risk receivers. The sectors having interwoven economic connections are vulnerable to each other, for example, the pair of bank and non-bank finance and the pair of mechanical equipment and building materials. There may be direct risk transmitting paths between sectors having service links, for instance, the finance sectors only receive risks from the auto sector because the finance sectors may provide loan to auto companies and to car buyers.

Then, a dynamic analysis is performed by means of a rolling window approach. It is observed that the connectedness peaks are associated with the financial crisis in 2008, ``money shortage'' in June 2013, the events of thousands of stocks slumped by the maximum 10\% daily limits in June 2015 and June 2018. More importantly, differing from the results of whole sample analysis, the sector of real estate acts as a role of risk receiver which is in contrast to our intention. This can be explained by that the real estate is the pillar industry of the Chinese economy and receives products and services from many sectors in its industry chain. It is also observed that the financial sectors play a critical important role in stabilizing the Chinese economic system during the turbulent periods. The recent trade war between the US and China also has significant impacts on the sectors of communication and computer, which turn from risk receivers to risk contributors. Furthermore, a risk transmitting path, mechanical equipment $\rightarrow$ chemical sector $\rightarrow$ transportation $\rightarrow$ sectors of bank, household appliances, and food \& drink, is observed.

Finally, the robustness of the model is tested with different parameters (lag order $p$, predictive horizon $H$, and rolling window width $W$). We find that the generalized variance decomposition method based on the VAR model and the rolling window method are not sensitive to the changes of model parameters, implying that our results are solid and sound. Our results not only uncover the spillover effects between the Chinese sectors, but also highlight the deep understanding of the patterns of risk contagion in the Chinese stock markets.

\section*{Acknowledgements}

We are grateful to Peng Wang, Mu-Yao Li, and Yin-Jie Ma for fruitful discussions. This work was partly supported by the National Natural Science Foundation of China (U1811462, 71532009, 91746108, 71871088), the Shanghai Philosophy and Social Science Fund Project (2017BJB006), the Program of Shanghai Young Top-notch Talent (2018), the Shanghai Outstanding Academic Leaders Plan, and the Fundamental Research Funds for the Central Universities.

\bibliographystyle{elsarticle-harv}
\bibliography{mybib}


\end{document}